\documentclass[twocolumn,prl,superscriptaddress,longbibliography,floatfix]{revtex4-1}

\usepackage{graphicx}
\usepackage{amsmath}
\usepackage{amssymb}
\usepackage{bm}
\usepackage{times}

\usepackage[pdftex]{hyperref}
\hypersetup{colorlinks=true,linkcolor=blue,citecolor=blue,urlcolor=blue}

\newcommand {\be}{\begin{equation}}
\newcommand {\ee} {\end{equation}}
\newcommand {\bea}{\begin{eqnarray}}
\newcommand {\eea} {\end{eqnarray}}
\newcommand{\non}{\nonumber}

\newcommand{\half}{{\textstyle\frac{1}{2}}}
\newcommand{\eps}{{\varepsilon}}

\begin{document}

\title{Multicritical point on the de Almeida-Thouless line in spin glasses in $d>6$ dimensions}

\author{M.A. Moore}
\affiliation{School of
Physics and Astronomy, University of Manchester, Manchester M13 9PL, UK}
\author{N. Read}
\affiliation{Department of Physics, Yale University, P.O. Box 208120, New Haven, CT 06520-8120, USA}
\date{January 29, 2018}
\begin{abstract}

The de Almeida-Thouless (AT) line in Ising spin glasses is the phase boundary in the temperature $T$ and
magnetic field $h$ plane below which replica symmetry is broken. Using perturbative renormalization group
(RG) methods, we show that when the dimension $d$ of space is just above $6$ there is a multicritical point
(MCP) on the AT line, which separates a low-field regime, in which the critical exponents have mean-field
values, from a high-field regime where the RG flows run away to infinite coupling strength; as $d$
approaches $6$ from above, the MCP approaches the zero-field critical point exponentially
in $1/(d-6)$. Thus on the AT line perturbation theory for the critical properties breaks down at
sufficiently large magnetic field even above $6$ dimensions, as well as for all non-zero fields when
$d\leq 6$ as was known previously. We calculate the exponents at the MCP to first order
in $\eps=d-6>0$. The fate of the MCP as $d$ increases from just above 6 to infinity
is not known.

\end{abstract}

\maketitle

The nature of the ordered phase of spin glasses has been controversial for decades. When various standard
calculational methods are applied to it, the results are sometimes in conflict. The picture which derives
from mean-field theory (valid at least for infinite dimensional systems) is that of replica
symmetry breaking (RSB) \cite{parisi:79,parisi:83,rammal:86,mezard:87,parisi:08}. However, the results of
real-space renormalization group (RG) calculations favor an ordered phase with replica symmetry when
the dimension $d$ of space is small \cite{moore:98,wang:17,wang:17c,angelini:15,angelini:17,angelini:17a}.
Recent calculations using the strong-disorder renormalization group were interpreted as suggesting that
the spin glass (SG) phase is replica symmetric for $d \le 6$ \cite{wang:17,wang:17c}.
Much of the debate on the existence or not of RSB has focussed on the de Almeida-Thouless (AT) line
\cite{dealmeida:78}. According to the RSB theory there is a phase transition in an applied magnetic field
$h$, occurring along the AT line $T_c(h)$ as the temperature $T$ is reduced. Below $T_c(h)$ there is the SG
phase with RSB, whereas for $T\geq T_c(h)$ replica symmetry is unbroken. The
existence of the AT line in high dimensions $d\geq 6$ is supported by, for example, Ref.\ \cite{singh:17}.
The existence of such a line in three dimensions has been the subject of  experimental work
\cite{mattsson:95} and controversial simulational studies \cite{joerg:08a,banos:12-ea,baity:14,
baityjanus:14}.

In early work, Bray and Roberts (BR) \cite{bray:80} derived a ``reduced'' field theory of
Landau-Ginzburg-Wilson type for a set of fluctuating fields that remain critical on the AT line.
Applying standard perturbative renormalization group (RG) methods at one-loop (i.e.\
lowest non-trivial) order, they showed that, when
$d$ is less than or equal to $6$, the coupling constants run away to infinity, so no stable physical RG
fixed point exists, and hence corrections to the mean-field exponents could not be calculated even at
leading order in $6-d$. (This is in contrast with the transition at $h=0$, for which such an expansion
exists in the conventional way \cite{harris:76}, using the unreduced theory.)
BR suggested that,
for $d<6$, the transition on the AT line could become first order, or the line itself could disappear.
When $d>6$, the BR RG flows have a domain of attraction of the zero coupling fixed point \cite{moore:11},
so that sufficiently small initial values of the couplings run towards zero, implying mean-field values
for critical exponents, while initial values outside this domain run off to infinity; this domain shrinks
to zero size as $d\to6^+$. It is also known that the form of the AT line at small $h$ is modified from
the mean-field result for dimensions $6<d<8$ \cite{green:83,fisher:85}.

\begin{figure}
\includegraphics[width=0.75\columnwidth]{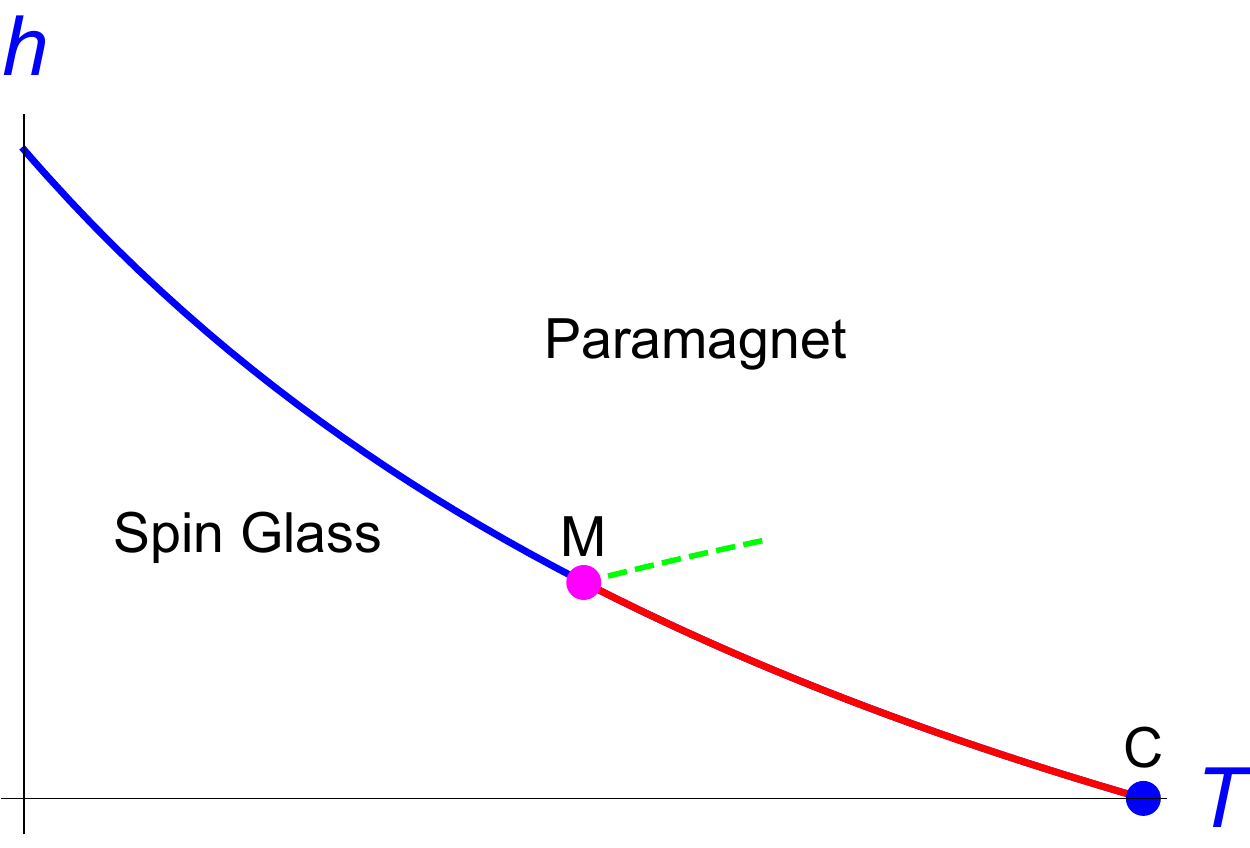}
\caption{(Color online) A schematic phase diagram, for dimension $d$ slightly larger than $6$, showing
the MCP $M$ on the AT line, at which the nature of the criticality changes.
The portion of the AT line at low $h$ (red) is where the exponents at the line are mean-field like;
the portion of the line at high $h$ (blue) is where the RG flows run away to infinity. The dashed
line (green) indicates schematically the direction along which the distinct correlation length exponent
of the multicritical fixed point might be observable. $M$ approaches the zero-field critical point $C$
as $d$ tends to $6$ from above.}
\label{mcp}
\end{figure}

In this paper we approach the problem from the point of view of dimensions $d$ larger than $6$. We calculate
the crossover from the unreduced to the reduced theory using perturbative RG methods at one-loop order.
We find that for sufficiently small $h$, the initial values of the couplings
in the BR theory lie inside the domain of attraction of zero coupling, but as $h$ increases they pass
through the boundary of the domain, and so run off to infinity. Hence there is a transition associated
with an RG fixed point on the boundary of the domain of attraction.
This implies that there is a multicritical point (MCP) $M$ at $(T,h)=(T_M,h_M)$ on the AT line for $d>6$,
at least for $d$ not much larger than $6$. For small fields, the critical behavior is that of mean field
theory, while for larger fields it is some other
unknown behavior (possibly first order); see Fig.~\ref{mcp}. The distance in temperature of the
MCP from the $h=0$ critical point varies as $c^{1/\eps}\to0$ as $\eps=d-6\to0^+$, where $c$ is some
constant ($0<c<1$). We calculate the exponents at the MCP at first order in $\eps$.

As a consequence, non-mean-field behavior of the critical properties occurs on a portion of the AT line
already for $d>6$. Its existence suggests the possibility of similar behavior for $d\leq 6$ as well, in this
case for all $h\neq0$. However, the possibility that the entire (non-mean-field-like) AT line disappears
at once for $d\leq 6$ cannot be excluded using the present methods. Clearly it is imperative to understand
the nature of the non-mean-field part of the AT line. The AT line is expected to intersect the
$T=0$ axis at $h=h_c$ when $d$ is finite. As $d\to\infty$ at fixed $T$ and $h$,
one expects that mean field theory becomes exact for the phase boundary and exponents, and hence that
both $h_c$ and $h_M$ should tend to infinity as $d$ increases; $M$ will reach $T=0$
either at some finite $d=d_u>6$, so there is no non-mean-field portion for $d> d_u$, or at $d_u=\infty$.

We start from the Edwards-Anderson (EA) model \cite{edwards:75} defined on a
$d$-dimensional hypercubic lattice of linear extent $L$ by the Hamiltonian
\begin{equation}
H = - \sum_{\langle ij \rangle} J_{ij} S_i S_j-h\sum_i S_i,
\label{eq:ham}
\end{equation}
where the summation is over distinct nearest-neighbor pairs only, the Ising spins take the values $S_i
\in \{\pm 1\}$ with $i = 1,2, \ldots, L^d$, and the random bonds $J_{ij}$ are independent Gaussian
variables of variance $(2d)^{-1/2}$ (chosen so that $T_c(h=0)\to1$ as $d\to\infty$) and zero
mean. From the partition function associated with
Eq.~(\ref{eq:ham}) one can derive \cite{harris:76,pytte:79,bray:79} the  replicated
and bond-averaged Landau-Ginzburg-Wilson field theory, which involves fluctuating fields
$Q_{\alpha \beta}=Q_{\beta\alpha}$, where as usual the indices  $\alpha$ and $\beta$ run over
values $1, 2, \cdots, n$, $n$ is set zero at the end of the calculation, and $Q_{\alpha\alpha}=0$ for all
$\alpha$. The action in this theory is \cite{bray:80}
\begin{eqnarray}
F[\{Q_{\alpha\beta}\}]  &=&   \int   d^dx\,   \left[- {\textstyle\frac{1}{4}}r \sum Q_{\alpha\beta}^2
+{\textstyle\frac{1}{4}}\sum(\nabla Q_{\alpha\beta})^2\right. \nonumber  \\
&&{} -   {\textstyle\frac{1}{6}}w\sum Q_{\alpha\beta}Q_{\beta\gamma}Q_{\gamma\alpha}
- {\textstyle\frac{1}{8}}y\sum Q_{\alpha\beta}^4\non\\
&&\left.\vphantom{\sum} -\half h^2
\sum Q_{\alpha\beta} + \ldots \right].
\label{FQ}
\end{eqnarray}
Here the summations in each term are over all values of the free indices in that term, and are
unrestricted except that $Q_{\alpha\alpha}=0$. Terms omitted are other less-important terms of order $Q^4$
or higher, or with more than two derivatives. The coefficients $w$ and $y$ are positive, while we have
reversed the usual sign of $r$, so that $r\propto T_c(h=0)-T>0$ for $T<T_c(h=0)$. This theory, to which
we refer as the unreduced theory, is usually believed to capture the essence of SG behavior near
criticality in $d$ dimensions.

The unreduced theory contains $\half n(n-1)$ modes when expanded to quadratic order, which can be
classified \cite{dealmeida:78} into symmetry types, conventionally called longitudinal (one mode),
anomalous ($n-1$ modes),
and replicon [$\half n(n-3)$ modes]. By a standard RG method, in which a cutoff of $1$ is
assumed, and Fourier components of fields with wavevectors in a shell just below the cutoff are successively
integrated out, followed by rescaling to restore the cutoff to $1$, one obtains the one-loop RG flow
equations
\cite{harris:76,fisher:85,pimentel:02} for the effective couplings $w(l)$, $r(l)$, $h(l)^2$, and $y(l)$
at length scale $e^l$ (where scale $l=0$ corresponds to the initial cutoff scale):
\bea
\frac{dw}{dl}&=&\half[-\varepsilon-3\eta]w-2w^3,\\
\frac{dr}{dl}&=&[2-\eta-4w^2]r,\\
\frac{dh^2}{dl}&=&\half[d+2-\eta]h^2,\\
\frac{dy}{dl}&=&[4-d-2\eta - Bw^2]y+Aw^4,
\eea
where $\eps=d-6$, $\eta=-\frac{2}{3}w^2$, and $A>0$ and $B$ are constants, the values of which are
not important. We adopted the convention of absorbing the geometric
factor $K_d=2/(\Gamma(d/2)(4 \pi)^{d/2})$ into $w^2$.
Mass corrections in denominators in these equations \cite{pimentel:02} have been dropped, except
in the RG equation for $r$ where the first order term has been retained. (That equation
should also include an inhomogeneous term that describes a shift in the critical temperature, however
that effect is also negligible in the limit we consider.)

The flow equations can be solved exactly. First, one has \cite{moore:11}
\be
w(l) =\frac{w_0e^{-\frac{1}{2}\eps l}}{\left[1+\frac{2w_0^2}{\varepsilon}(1-e^{-\eps
l})\right]^{1/2}},
\ee
where $w_0=w(0)$, which typically is of order 1. We will see that the matching to the BR reduced action
that we require occurs at
$w\propto\sqrt{\eps}$ as $\eps\to0^+$ (i.e.\ $d\to 6^+$), so that the limit of interest in the following
is always $\eps\to0$ with $\eps l$ (and $w_0$) fixed. In this limit,
\be
w(l) =\left(\frac{\eps}{2}\right)^{1/2}\frac{e^{-\frac{1}{2}\eps l}}{(1-e^{-\eps l})^{1/2}}\left[1
+{\cal O}(\eps)\right].
\ee
(Recall that $f={\cal O}(g)$ as $\eps\to0$ means that $|f(\eps)/g(\eps)|$ is bounded above for all
$\eps$ sufficiently close to $0$.) Similarly \cite{moore:11},
\bea
r(l)&=&r(0)\exp\left[2l-{\textstyle\frac{10}{3}}\Delta(l)\right],\\
h(l)^2&=&h(0)^2\exp\left[\half(d+2)l+{\textstyle\frac{1}{3}}\Delta(l)\right],
\eea
where
\bea
\Delta(l) &=& \int_0^lw(l')^2\,dl'\\
          &=&\half\ln\left[1+\frac{2w_0^2}{\eps}(1-e^{-\eps l})\right]\\
          &=&\half\ln\left[\frac{2w_0^2}{\eps}(1-e^{-\eps l})\right]+{\cal O}(\eps)
\eea
in the required limit. For $y$, we obtain likewise
\bea
\lefteqn{y(l)=y(0)\exp\left[(4-d)l+({\textstyle\frac{4}{3}}-B)\Delta(l)\right]}&&\\
&&{}+Ae^{(4-d)l+(\frac{4}{3}-B)\Delta(l)}\int_0^lw(l')^4e^{-(4-d)l'-(\frac{4}{3}-B)\Delta(l')}\,dl'.\non
\eea
In this case the required limit can be obtained by defining the integration variable $l''=\eps l'$, in
terms of which the integration limit becomes a constant and Laplace's method can be
applied to the integral, to obtain
\be
y(l)=A\frac{\eps^2 e^{-2\eps l}}{8(1-e^{-\eps l})^2}\left(1+{\cal O}(\eps)\right)
\ee
as $\eps\to0$ with $\eps l>0$ fixed; the initial value $y(0)$ is an exponentially small correction and
has been dropped. Thus $y=\half A w^4$ \cite{fisher:85}.

The crossover to the BR reduced action takes place at the scale $l=l^\ast$ at which the longitudinal
and anomalous modes have mass-squared $1$, while by definition of the AT line, the replicons
remain massless there. The action (\ref{FQ}) predicts at mean-field level that for non-zero $h^2$,
the AT line in the $r$--$h$ plane and the replica symmetric expectation $Q$ of $Q_{\alpha\beta}$ on the
line are given by
\be
Q=\frac{r}{2w}>0,\quad h^2=2yQ^3=\frac{yr^3}{4w^3}.
\ee
These expressions \cite{bray:80} are valid up to corrections of
relative size $yr/w^2$. Further, the longitudinal and anomalous modes can be shown \cite{dealmeida:78}
to have mass-squared
$r$ on the AT line. Setting $r(l^*)=1$, the corrections to the leading expressions for $Q$ and
$h^2$ on the AT line are of relative order $y(l^*)r(l^*)/w(l^*)^2=\half A w(l^*)^2={\cal O}(\eps)$
as $\eps\to0$ with $\eps l^*$ fixed.

The BR reduced action results from (\ref{FQ}) by setting the fluctuations
of the non-replicon modes to zero; the fields $\widetilde{Q}_{\alpha\beta}$ in the replicon sector are
defined by the condition $\sum_{\beta}\widetilde{Q}_{\alpha \beta}=0$ for all $\alpha$, in addition to
$\widetilde{Q}_{\alpha\alpha}=0$. The reduced action is \cite{bray:80}
\begin{eqnarray}
F[\{\widetilde{Q}_{\alpha\beta}\}] &=&  \int   d^dx\,   \left[{\textstyle\frac{1}{4}}\widetilde{r}
\sum \widetilde{Q}_{\alpha\beta}^2
+{\textstyle\frac{1}{4}}\sum(\nabla
\widetilde{Q}_{\alpha\beta})^2 \right.   \\
&&\left.{}-{\textstyle\frac{1}{6}}w_1
\sum \widetilde{Q}_{\alpha\beta}
\widetilde{Q}_{\beta\gamma}\widetilde{Q}_{\gamma\alpha}
-{\textstyle\frac{1}{6}}w_2\sum \widetilde{Q}_{\alpha\beta}^3\right],\nonumber
\label{FBR}
\end{eqnarray}
up to terms higher order in $\widetilde{Q}$ or derivatives.
Here again the summations are unrestricted, but the fields obey the conditions noted above.
$\widetilde{r}=-r+2wQ$ \cite{bray:80} vanishes on the AT line. In principle the non-replicon modes should be
integrated out exactly once they become massive [$r(l)\geq 1$], not just projected to zero, but this
should produce at most only negligible [${\cal O}(\eps)$] corrections to coefficients, because (as we will
see) the couplings are of order $\eps^{1/2}$.

The remaining coupling constants in the BR reduced action are
\bea
w_1&=&w-3uQ,\\
w_2&=&3yQ.
\eea
Here $u$ is another quartic coupling in the unreduced action, for which the flow is the same as
for $y$ except that $A$ and $B$ are replaced by some $A'$ and $B'$. By the above, when the crossover
to the BR reduced action occurs, these are
\bea
w_1&=&w,\\
w_2&=&0,
\eea
plus terms of order ${\cal O}(\eps w)$ in the required limit.

When $w_2=0$, the critical $w_1$ on the boundary of the domain of attraction of the origin in the BR RG
flows (which we review below) is $w_1=(c'\eps)^{1/2}$, where $c'=1/24$. Setting $w(l_M^*)=(c'\eps)^{1/2}$
gives for $l^*=l_M^*$ at $M$
\be
\frac{e^{-\eps l_M^*}}{1-e^{-\eps l_M^*}}=2c',
\ee
that is,
\be
e^{-\eps l_M^*} =\frac{2c'}{1+2c'}<1.
\ee
Thus $\eps l_M^*>0$ is a constant, and $l_M^*$ is large as $\eps\to0$, regardless of the
precise value of $c'$ (note that $c'>0$).

Using the expressions for $r(l^*)=1$ and $h(l^*)^2$, we find the location of the MCP $M$ in terms of the
bare (i.e.\ lattice scale, $l=0$) parameters in the unreduced action,
\bea
r(0)_M&=& \eps^{-5/3}\left( e^{-\eps l_M^*}\right)^{2/\eps}\left[2w_0^2(1-e^{-\eps l_M^*})\right]^{5/3},\\
h(0)_M^2&=&\frac{A\eps^4 r(0)_M^2 w_0 e^{-\eps l_M^*}}{8\left[2w_0^2(1-e^{-\eps l_M^*})\right]^4},
\eea
in the limit as $\eps\to0$, where $\eps l_M^*$ and $w_0$ are constant.
These are among the main results of this paper; they show that the MCP $M$ approaches the
critical point $C$ exponentially fast as $\eps\to0$.
The exponent in $\eps^{-5/3}$ in the first formula should be universal.
In the second formula, the fact that $h(0)\propto r(0)$ as $d\to6$ (neglecting the prefactor)
agrees with Refs.\ \cite{green:83,fisher:85} (it was derived in a similar way in Ref.\
\cite{fisher:85}), while the $\eps^4$ in the coefficient agrees with the results of Ref.\ \cite{moore:11};
note however that the results of these references were valid in the different limit $r(0)\to0$ at fixed
$\eps$, followed by the $\eps\to0$ limit \cite{parisi:12}.

Next we turn to calculations that make greater use of the BR reduced theory.
BR obtained the one-loop RG equations \cite{bray:80,pimentel:02}:
\bea
\frac{dw_1}{dl} & = & \half
\left[-\eps -3\tilde{\eta}\right]w_1+14w_1^3-36w_1^2w_2\non\\
&&{}+18w_1w_2^2+w_2^3,
\label{ATRG1}\\
\frac{dw_2}{dl} & = & \half\left[-\eps -3\tilde{\eta}\right]
w_2+24w_1^2w_2\non\\
&&{}-60w_1w_2^2+34w_2^3,
\label{ATRG2}\\
\frac{d \tilde{r}}{d l}&=& [2-\tilde{\eta}] \tilde{r}-\frac{3 \tilde{\eta}}{(1+\tilde{r})^2},
\label{ATRGr}
\eea
where now $\tilde{\eta}=(4w_1^2-16w_1w_2+11w_2^2)/3$.

\begin{figure}
\includegraphics[width=0.75\columnwidth]{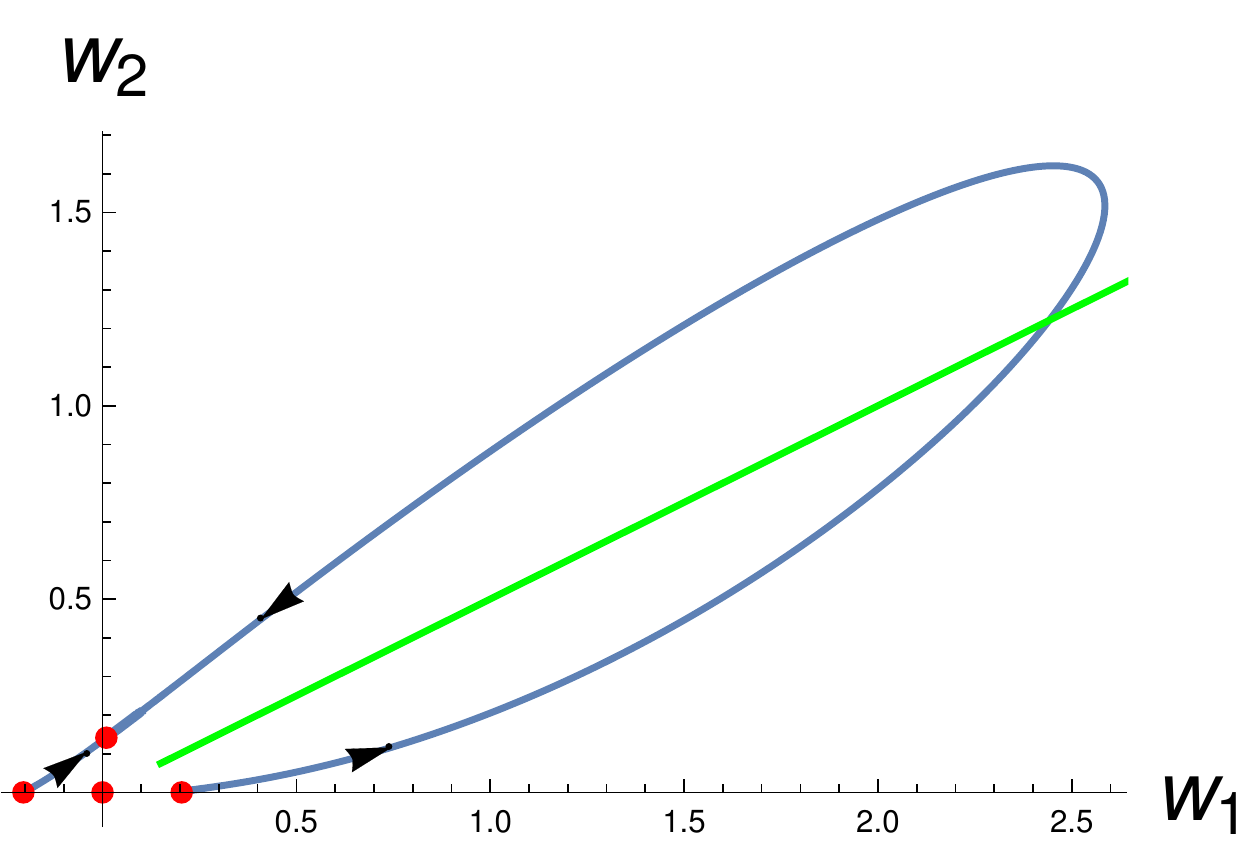}
\caption{(Color online) The domain of attraction of the Gaussian fixed
point $w_1=w_2=0$ of the BR RG flow equations for $d>6$ is bounded by the separatrix shown
(blue curve). Distances are measured in units of $\sqrt{\epsilon}$.
Only the region $w_2>0$ is displayed; the other part is obtained by inversion symmetry. RG fixed points
are shown as dots (red). The straight line (green) indicates the initial values.}
\label{fig:domain}
\end{figure}

Eqs.~({\ref{ATRG1})-(\ref{ATRG2}) were solved numerically for $d>6$ in Ref.\ \cite{moore:11}. In the
$w_1$--$w_2$ plane, the Gaussian fixed point $G$ at $w_1=w_2=0$ is stable for $d>6$. $G$ is the attractor
for flows inside the domain of attraction, as shown in Fig.\ \ref{fig:domain}. The boundary of the domain
is itself a flow line of the RG: a separatrix. There are two pairs of other fixed points on the separatrix
\cite{bray:80};
the fixed points $U$ and $-U$ at $w_1=\pm \sqrt{\eps/24}$, $w_2=0$, which are unstable, and the fixed
points $Z$ and $-Z$ at $(w_1,w_2)= \pm\sqrt{\eps}(0.00983702, 0.141449 )$, which have one stable (incoming)
direction along the separatrix, and one unstable direction (marked in orange in Fig.~\ref{fig:flows}).
Outside the separatrix, all flows go to infinity.

The exponents for the MCP can be obtained by standard methods from the RG equations linearized at a fixed
point. Although we
found above that the initial values for the BR flows cross the separatrix at a point approaching $w_2=0$
as $\eps\to0$, the generic case for $\eps>0$ does not pass through that point, and consequently the fixed
point that controls the true asymptotics of the MCP is $Z$ (see Fig.~\ref{fig:flows}).  First, the
exponent $\eta$ that describes the power law decay of the replicon correlation function on the AT line,
\begin{equation}
\overline{(\langle S_i S_j \rangle-\langle S_i \rangle \langle S_j \rangle)^2} \sim \frac{1}{r_{ij}^{d-2
+\eta}}
\label{criteta}
\end{equation}
(where the overline represents the average over the $J_{ij}$), can be found by evaluating $\tilde{\eta}$
at $Z$, giving $\eta=0.06607 \eps$ at $M$. Next, the BR RG equations for $w_1$ and $w_2$, when linearized
about $Z$, produce the eigenvalues $\eps$ for the unstable direction, and $-0.25624 \cdots \eps$ for the
stable direction along the separatrix. The first of these describes the crossover as
the system is perturbed off the MCP but staying on the AT line, flowing to $G$ if $h^2$ is decreased.
(The second gives corrections to scaling.) Finally, by linearizing Eq.\ (\ref{ATRGr}) about the fixed
point value of $\tilde{r}$, one can calculate the exponent $\nu$ for the correlation length as the AT line
is approached; it is given by $1/\nu=2 +5 \tilde{\eta}$, so at $Z$ (i.e.\ $M$), $\nu =1/2-0.082585 \eps$.
Hyperscaling relations among exponents are satisfied at $M$, even though $d>6$. The mean-field portion
of the AT line at low $h$ is governed instead by $G$, with $\eta=0$ and $\nu=1/2$;
hyperscaling is violated for $d>6$.

\begin{figure}
\includegraphics[width=0.75\columnwidth]{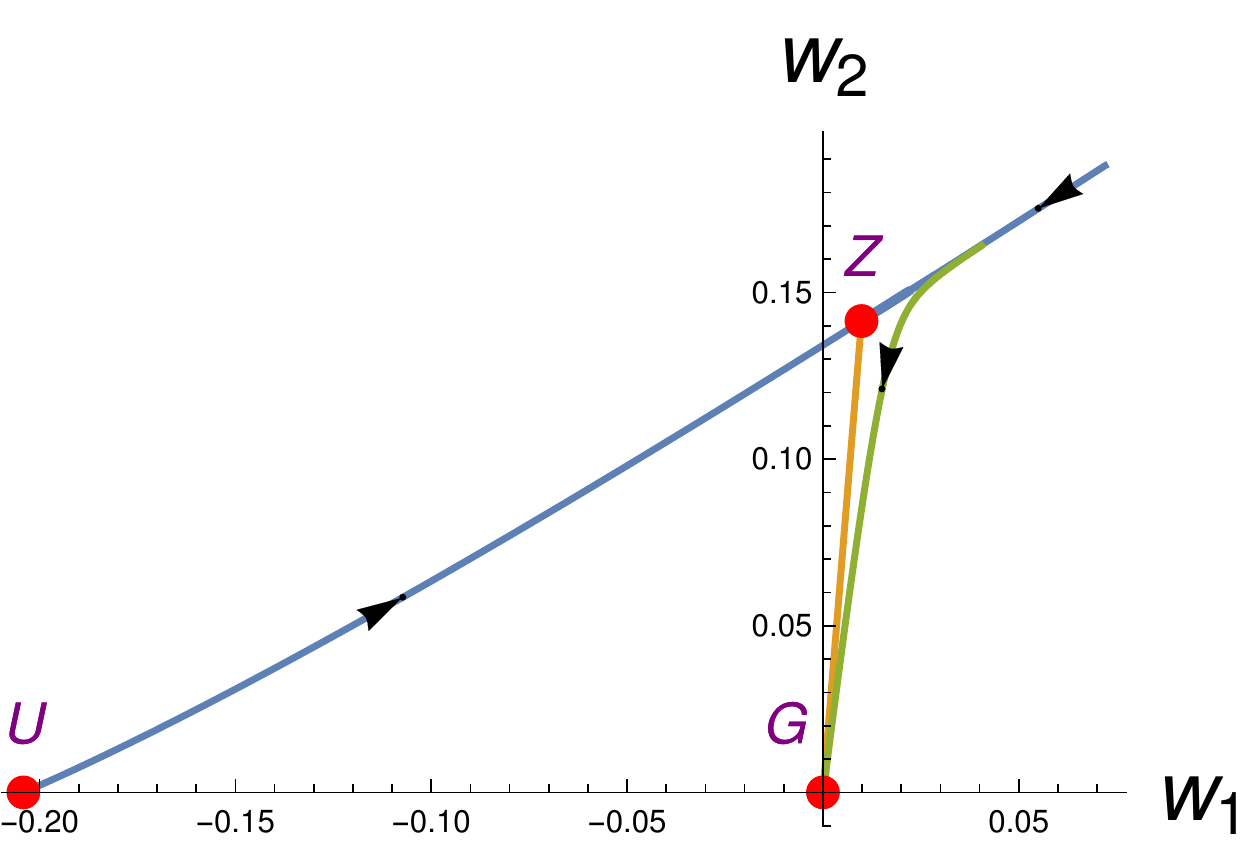}
\caption{(Color online) An enlarged version of Fig.~\ref{fig:domain} in the vicinity of the origin.
The orange line connects the fixed point $Z$ to the Gaussian fixed point $G$ at the origin. The green curve
is a trajectory for an initial point just off the separatrix.}
\label{fig:flows}
\end{figure}

We emphasize that our results are well controlled within perturbation theory, similarly to the usual
$\eps$ expansion. They predict a breakdown of perturbation theory on the AT line above a critical $h$
when $d>6$. Hence the only way they could fail to be correct, or the MCP not exist, would be if
perturbation theory broke down at all $h>0$ when $d$ is just larger than $6$; it is unclear how that would
occur. Similar methods show that the one-dimensional power-law model \cite{kotliar:83} possesses a MCP
in the region corresponding to $d>6$.

It is not clear what happens to the MCP as $d$ increases further. The transition point $h=h_c$ at $T=0$
is expected
to go to infinity as $d\to\infty$. In finite dimensions, the Bethe-Peierls approximation, that is, the
solution of the SG on the Bethe lattice \cite{bowman:82,thouless:86}, predicts a transition at
$T_c(h)\to0$ at non-zero $h=h_c<\infty$,
and $h_c^2\sim \ln d\to\infty$ as $d\to\infty$ \cite{parisi:14}. (For Gaussian random fields with
variance $h^2$ and mean zero, $h_c\sim \sqrt{2d}$ instead.) This limit thus agrees with the solution of the
Sherrington-Kirkpatrick model \cite{dealmeida:78}. On the Bethe
lattice, the $T=0$ transition is percolative in nature \cite{doty:89}, but it is not clear if that is
true for the EA model at high $d$. No MCP has been found in these other models, so we expect
$T_M\to0$ as $d\to\infty$. Hence we define a dimension $d_u>6$ at which
the MCP hits the $T=0$ $h$-axis; possibly, $d_u=\infty$.

Recent work has suggested alternative pictures. Ref.\ \cite{angelini:15} finds a non-mean-field transition
governed by a zero-temperature fixed point at non-zero $h$ for sufficiently high dimensions, and no MCP
(see also Ref.\ \cite{bray:84}).
However, the exponents for that transition given there for hierarchical-lattice models imply that the SG
susceptibility exponent $\gamma$ is negative, which means the SG susceptibility does not diverge at the
AT line, at variance with the conventional view of the line. In those models any finite region
is contained in a region with only two spins on the boundary, so there can be at most four ground states.
This and a similar limitation on the number of pure states at $T>0$ preclude most forms of RSB {\it a
priori} (see also Ref.\ \cite{gertler:17}). In other work, Ref.\ \cite{charbonneau:17} has extended the
BR calculation to three-loop order and suggested that a fixed point might be present at strong coupling
in 5 dimensions and below, even if not right up to 6 dimensions. Their argument is of unknown validity.

In conclusion, we have shown that there is a non-perturbative (non-mean-field) portion
of the AT line in a spin glass in dimensions greater than $6$, separated from the mean-field region at
low magnetic field by a multicritical point. This suggests, though it does not prove, that a similar
non-perturbative AT line could also persist below six dimensions.

\begin{acknowledgments}

One of us (M. A. M) would like to thank Maria Chiara Angelini,  Giulio Biroli, Gilles Tarjus, Tamas
Temesv\'{a}ri, and Sho Yaida for email discussions and Mike Godfrey for his help with Mathematica.

\end{acknowledgments}

\bibliography{mcp}

\end{document}